\documentclass[aps,psfig,epsfig,preprint] {revtex4}
\usepackage{epsfig}
\usepackage{color}
\usepackage{bm}
\usepackage{subfigure}

\begin{document}
\draft
\title{
{\normalsize \hskip4.2in USTC-ICTS-07-12} \\{\bf Signals of
Unparticles in Low Energy Parity Violation and NuTeV Experiment}}

\author{Gui-Jun Ding\footnote{E-mail: dinggj@ustc.edu.cn}, Mu-Lin
Yan\footnote{E-mail: mlyan@ustc.edu.cn}}

\affiliation{\centerline{Interdisciplinary Center for Theoretical
Study and Department of Modern Physics,} \centerline{University of
Science and Technology of China, Hefei, Anhui 230026, China} }

\begin{abstract}

We have studied the possible signals of unparticle in atomic parity
violation(APV) along an isotope chain and in the NuTeV experiment.
The effects of unparticle physics could be observed in APV, if the
uncertainty in relative neutron/proton radius shift
$\delta(\Delta\frac{R_N}{R_P})$ is less than a few times $10^{-4}$
by measuring the parity violating electron scattering. The
constraints imposed by NuTeV experiment on unparticle physics are
discussed in detail. If the NuTeV results are confirmed by future
experiments, we suggest that unparticle could account for a part of
NuTeV anomaly. There exist certain regions for the unparticle
parameters ($\Lambda_{\cal U}$, $d_{\cal U}$, $c_{V{\cal U}}$ and
$c_{A{\cal U}}$), where the NuTeV discrepancy could be completely
explained by unparticle effects and the strange quark asymmetry,
even with or without the contributions from the isoscalarity
violation etc. It is remarkable that these parameter regions are
consistent with the constraints from $b\rightarrow s\gamma$

\vskip 0.5cm

\end{abstract}
\maketitle
\section{introduction}
Recently H.Georgi suggests that the scale invariant stuff may exist
in our world, which is made of unparticles\cite{georgi1,georgi2}.
The unparticle weakly couples to the Standard Model(SM) matter,
which is suppressed by the inverse of large energy scale from the
view of the effective theory. This scenario serves as a possible
component of the physics beyond Standard Model above TeV scale,
which can be explored experimentally at the LHC and ILC in future.
Although conformal symemetry is widely and deeply studied in the
condense matter physics (mostly in 2-dimension) and the superstring
theory (generally in high dimension and with the supersymmetry ), we
know little about the conformal symmetry in 4-dimension and how it
is manifested in particle physics experiments, if unparticle really
exists.

Unparticle is very different from the ordinary matters, some exotic
and remarkable properties have been demonstrated, ${\it i.e.,}$ it
looks like a non-integral number $d_{\cal U}$ of invisible massless
particles with unusual phase space, where $d_{\cal U}$ is the scale
dimension of the relevant unparticle operator ${\cal O}_{\cal{U}}$.
The peculiar two-point correlation function of unparticle can
produce unusual interference patterns in the time-like region and so
on\cite{georgi1,georgi2}. The phenomenological consequences of
unparticle and its possible signatures have been discussed by
several authors. In a previous paper\cite{ding}, we have discussed
the virtual effects of unparticle in the DIS processes to the lowest
nontrivial order of the couplings of the unparticle with the SM
fields. The leading order corrections to the structure functions are
the interference terms between the vector unparticle exchange
amplitudes and the standard model amplitudes in the space-like
region\cite{ding}. Unparticle production in $Z\rightarrow
f\overline{f}\cal{U}$ and $e^{+}e^{-}\rightarrow\gamma\cal{U}$,
unparticle effects in the Drell-Yan process, fermionic unparticle,
its physical effects on muon anomalous magnetic moments,
$B^{0}-\overline{B}^{0}$ mixing and $D^{0}-\overline{D}^{0}$ mixing,
the lepton flavor violating processes, as well as the bounds on
unparticle from the Higgs sector and the deconstruction of
unparticle {\it etc} have also been investigated
\cite{yuan,luo,geng,liao,aliev,li,duraisamy,lu,stephanov,fox,greiner,
davou,dgm,he,acg,indian,Ding:2007jr}.

As we have shown in Ref.\cite{ding}, the low energy effective
interactions between the SM fermions and unparticle can take the
following form:
\begin{equation}
\label{2}\frac{1}{\Lambda_{\cal U}^{d_{\cal
U}-1}}\overline{f}\gamma_{\mu}(c_{V{\cal U}}+c_{A{\cal
U}}\gamma_5)f\;{\cal O}^{\mu}_{\cal{U}}\;,~~~\frac{1}{\Lambda_{\cal
U}^{d_{\cal U}-1}}\overline{f}(c_{S{\cal U}}+ic_{P{\cal
U}}\gamma_5)f\;{\cal O}_{\cal{U}}
\end{equation}
where $d_{\cal U}$ is the scale dimension of the relevant unparticle
operator (${\cal O}^{\mu}_{\cal{U}}$ or ${\cal O}_{\cal{U}}$), and
${\cal O}^{\mu}_{\cal{U}}$ is a transverse and hermitian operator.
$c_{V{\cal U}}$, $c_{A{\cal U}}$, $c_{S{\cal U}}$ and $c_{P{\cal
U}}$ are dimensionless coupling constants. Those effective
interactions are the leading order interactions between the SM
fermions and the vector(scalar) unparticle, and universal couplings
have been assumed here.

Scale invariance determines the two-point correlation function of
unparticle operator to be
\begin{eqnarray}
\label{add1}\int d^4x\,e^{iP\cdot x}\langle0|{\cal O}_{\cal
U}(x)\,{\cal O}_{\cal U}^{\dagger}(0)|0\rangle&=&\frac{A_{d_{\cal
U}}}{2\,\sin(d_{\cal U}\pi)}\;\frac{i}{(-P^2-i\epsilon)^{2-d_{\cal
U}}}\\
\label{add2}\int d^4x\,e^{iP\cdot x}\langle0|{\cal O}^{\mu}_{\cal
U}(x)\,{\cal O}_{\cal U}^{\nu\dagger}(0)|0\rangle&=&\frac{A_{d_{\cal
U}}}{2\,\sin(d_{\cal
U}\pi)}\;\frac{i(-g^{\mu\nu}+P^{\mu}P^{\nu}/P^2)}{(-P^2-i\epsilon)^{2-d_{\cal
U}}}
\end{eqnarray}
where
\begin{equation}
\label{add3}A_{d_{\cal U}}=\frac{16\pi^{5/2}}{(2\pi)^{2d_{\cal
U}}}\frac{\Gamma(d_{\cal U}+\frac{1}{2})}{\Gamma(d_{\cal
U}-1)\Gamma(2d_{\cal U})}
\end{equation}

In this work, we shall study the particular aspects of unparticle
physics in atomic parity violation(APV) along an isotope chain and
in the NuTeV experiment. Since both APV and NuTeV experiment are
important low energy probes to new physics beyond SM, which
complements new physics searches at high energy colliders (such as
LHC and ILC). Therefore, they should play important role in
exploring unparticle physics together with LHC and ILC in future.
The paper is organized as follows. In section II, the physics
effects of unparticle in APV along an isotope chain and the
associated theoretical uncertainties are investigated. We
concentrate on the possible signals of unparticle in NuTeV
experiment in section III. Finally we present summary and our
conclusions.

\section{unparticle in low energy parity violation}

Low energy parity violation observables have played important role
in uncovering and testing the structure of the electroweak sector of
the standard model. They provide important information for physics
beyond SM, and place stringent constraints on a variety of new
physics scenarios. Low energy parity violation and collider
experiments(such as LHC and ILC) provide powerfully complementary
probes of new physics at the TeV scale.

The basic quantity of interest in considering weak neutral current
parity violation is the so-called weak charge $Q_W$. This quantity
is the weak neutral current analog of the electron-magnetic charge,
which characterizes the strength of the electron axial vector times
nucleus(or electron) vector weak neutral current interaction. The
most precise determination of $Q_W$ has been obtained with the
atomic parity violation in Cs(cesium) by the Boulder
group\cite{cesium}.

There are generally large theoretical uncertainties from atomic
structures for the APV observables of a single isotope. A strategy
for evading thess atomic structure uncertainties is to measure the
ratios of parity violation observables along an isotope
chain\cite{isotope}, and the isotopes of Ba(Barium) and
Yb(Ytterbium) are currently under study at Seattle and Berkeley
respectively. Two ratios are usually considered
\begin{eqnarray}
\label{add4}R_1&=&\frac{A^{NSID}_{PV}(N')-A^{NSID}_{PV}(N)}{A^{NSID}_{PV}(N')+A^{NSID}_{PV}(N)}\\
\label{add5}R_2&=&\frac{A^{NSID}_{PV}(N')}{A^{NSID}_{PV}(N)}
\end{eqnarray}
where $A^{NSID}_{PV}(N)$ is the nuclear spin-independent part of the
APV observable for an atom with neutron number $N$, which is given
by
\begin{equation}
\label{add6}A^{NSID}_{PV}(N)=\xi\,Q_{W}
\end{equation}
where $\xi$ contains all the atomic structure effects for a point
nucleus including the many-body corrections. For a single isotope, a
significant source of theoretical uncertainties arises from the
computation of the atomic structure dependent constant $\xi$. It was
shown in Ref.\cite{nuclstu} that taking ratios between isotopes
cancels essentially all dependence on the atomic structure and the
associated uncertainties. However, the nuclear structure dependence
does not cancel in the ratios, especially the ratios would receive
corrections from the neutron distribution. The dependence of the
parity violation amplitude on the nuclear structure can be
incorporated through the corrections to the weak charge, then
\begin{equation}
\label{rev1}Q_W=Q^0_W+Q^{nucl}_W+Q^{{\cal U}}_W
\end{equation}
where $Q^0_W$ is the nuclear weak charge in the standard model.
Including the effects of radiative corrections, $Q^0_W$
becomes\cite{Marciano:1990dp}
\begin{equation}
\label{v4eq1}Q^0_W=(0.9857\pm0.0004)(1+0.00782T)\{-N+Z[1-(4.012\pm0.010)\bar{x}]\}
\end{equation}
where $\bar{x}\equiv\sin^2\theta_{W}$. $\bar{x}$ is defined at the
mass scale $m_Z$ by the modified minimal substraction and is given
by\cite{Marciano:1990dp}
\begin{equation}
\label{v4eq2}\bar{x}=0.2323+0.00365S-0.00261T
\end{equation}
In Eq.(\ref{v4eq1}) and Eq.(\ref{v4eq2}), $S$ is the Peskin-Takeuchi
parameter characterizing the isospin-conserving quantum loop
corrections, and $T$ characterizes the isospin-breaking
corrections\cite{Peskin:1990zt}. $Q^{{\cal U}}_W$ is the correction
to the weak charge induced by unparticle, which can be
straightforwardly calculated\cite{Bhattacharyya:2007pi}
\begin{equation}
\label{add8} Q^{\cal U}_{W}=\frac{24}{\Lambda^2_{\cal
U}G_F}\frac{(2\pi)^{3/2-2d_{\cal U}}\Gamma(d_{\cal
U}+1/2)}{\Gamma(d_{\cal U})(2d_{\cal U}-1)}(Z+N)\,c_{V{\cal
U}}c_{A{\cal U}}
\end{equation}
The scalar and pseudoscalar couplings $c_{S{\cal U}}$, $c_{P{\cal
U}}$ between unparticle and SM fermions don't contribute to the weak
charge of nuclei. These two couplings are forbidden by the SM
symmetry if unparticle ${\cal O}_{\cal U}$ is SM singlet. The
nuclear structure corrections are contained in
$Q^{nucl}_W$\cite{isotope,nuclstu}, which is given by
\begin{equation}
\label{rev2}Q^{nucl}_W\simeq -N(q_n-1)+Z(1-4\bar{x})(q_p-1)
\end{equation}
where $q_p=\int\rho_p(r)f(r)d^3r$, $q_n=\int\rho_n(r)f(r)d^3r$,
$\rho_p(r)$ and $\rho_n(r)$ are respectively the proton and neutron
distributions within atomic nuclei, and $f(r)$ is the spatial
variation of the electron wavefunction inside the nucleus. In order
to get an idea of the relative importance of unparticle
contributions and nuclear structure effects, we follow
Ref.\cite{nuclstu} and consider a simple model in which the nucleus
is treated as a sphere of uniform proton and neutron number
densities out radii $R_P$ and $R_N$ respectively. In this case, one
obtains\cite{nuclstu}
\begin{eqnarray}
\nonumber&&q_n=1-(Z\alpha)^2g(r_N)+{\cal O}((Z\alpha)^4)\\
\label{v4eq3}&&q_p=1-0.260(Z\alpha)^2+{\cal O}((Z\alpha)^4)
\end{eqnarray}
where
\begin{equation}
\label{v4eq4}g(r_N)=\frac{3}{10}\,r^2_N-\frac{3}{70}\,r^4_N+\frac{1}{450}\,r^6_N\,,~~~~r_N=\frac{R_N}{R_P}
\end{equation}
The above estimate of $q_n$ yields the same general trend as the
detailed calculations from the Skyrme model, relativistic or
nonrelativistic Hartree-Fock nuclear calculations, and the mean
field methods etc, with the absolute values differing generally by
parts in a thousand or less\cite{nuclstu}. After including both the
unparticle and the nuclear structure corrections into the ratios of
APV, we have
\begin{eqnarray}
\label{add9}R_1&\simeq&\frac{Q_{W}(N')-Q_{W}(N)}{Q_{W}(N')+Q_{W}(N)}=R_{1}^{0}(1+\delta^{nucl}_1+\delta^{\cal U}_1)\\
\label{add10}R_2&\simeq&\frac{Q_{W}(N)}{Q_{W}(N)}=R^{0}_{2}(1+\delta^{nucl}_2+\delta^{\cal
U}_2)
\end{eqnarray}
where
\begin{eqnarray}
\label{add11}R^{0}_1&=&\frac{Q^{0}_{W}(N')-Q^{0}_{W}(N)}{Q^{0}_{W}(N')+Q^{0}_{W}(N)}\simeq\frac{\Delta
N}{N+N'-2(1-4.012\bar{x})Z}\\
\label{add12}R^{0}_2&=&\frac{Q^{0}_{W}(N')}{Q^{0}_{W}(N)}\simeq\frac{N'-(1-4.012\bar{x})Z}{N-(1-4.012\bar{x})Z}
\end{eqnarray}
which give the ratios in the standard model at the leading order.
$\delta^{nucl}_1$ and $\delta^{nucl}_2$ denote the nuclear structure
corrections to $R_1$ and $R_2$ respectively, $\delta^{\cal U}_1$ and
$\delta^{\cal U}_2$ are the corresponding corrections arising from
unparticle with $N'=N+\Delta N$. Dropping the terms containing the
factor $1-4.012\bar{x}$, which should be quite negligible due to the
accidental value of $\bar{x}\approx\frac{1}{4}$, then we obtain
\begin{eqnarray}
\label{rev3}&&\delta^{nucl}_1=-(Z\alpha)^2\frac{N'}{\Delta N}f'(r_{N})\Delta r_{N}\\
\label{add13}&&\delta^{\cal U}_1=\frac{24}{\Lambda^2_{\cal
U}\,G_F}\frac{(2\pi)^{3/2-2d_{\cal U}}\Gamma(d_{\cal
U}+1/2)}{\Gamma(d_{\cal U})(2d_{\cal
U}-1)}\frac{2Z}{N+N'}\,c_{V{\cal
U}}c_{A{\cal U}}\\
\label{rev4}&&\delta^{nucl}_2=-(Z\alpha)^2f'(r_N)\Delta
r_N\\
\label{add14}&&\delta^{\cal U}_2=\frac{24}{\Lambda^2_{\cal
U}\,G_F}\frac{(2\pi)^{3/2-2d_{\cal U}}\Gamma(d_{\cal
U}+1/2)}{\Gamma(d_{\cal U})(2d_{\cal U}-1)}\frac{Z}{N}\frac{\Delta N
}{N'}\,c_{V{\cal U}}c_{A{\cal U}}
\end{eqnarray}
where $\Delta r_N=(R_{N'}-R_{N})/R_P$, which denotes the shift in
the neutron radius(relative to the proton radius) along the isotope
chain. It is obvious that $\delta^{\cal U}_2$ contains an explicit
suppression factor $\Delta N/N'$, therefore $\delta^{\cal U}_1$ is
more sensitive to unparticle physics than $\delta^{\cal U}_2$ for a
given experimental precision. From Eq.(\ref{rev3})-Eq.(\ref{add14}),
we can see that both $\delta^{\cal U}_1$ and $\delta^{\cal U}_2$
depend on the product of $c_{V{\cal U}}$ and $c_{V{\cal U}}$, and
the dependence of $\delta^{nucl}_1$ on the variation in neutron
radius along the isotope is enhanced by a factor $N'/\Delta N$.
Thus, if one uses the ratios of APV in different isotopes to learn
about the new physics contributions from unparticle, one should have
extremely precise knowledge of the shift in neutron radius.

So far the proton distribution is well known from electric probes:
electron and muon scattering, optical isotope shifts, muonic atoms
etc. Whereas there exist no reliable experimental determinations of
the neutron distribution so far. Explicit studies of isotope shift
$\Delta r_{N}$ associated with $\rho_n(r)$ have been reported in
Refs.\cite{nuclstu,Chen:1993fw}, these authors showed that the
spread in the predictions of $\Delta r_{N}$ corresponds to a $100\%$
uncertainties in the model average for $\Delta r_{N}$. The detailed
knowledge of neutron distribution is crucial to APV. A
model-independent experimental determination of the neutron
distribution $\rho_n(r)$ can be achieved by parity violating
electron scattering\cite{Horowitz:1999fk}. A precise determination
of $\rho_n$ using parity violating electron scattering would
sufficiently constrain model calculations so as to significantly
reduce the theoretical isotope shift uncertainties.

\begin{figure}[hptb]
\begin{center}
\includegraphics*[width=10cm]{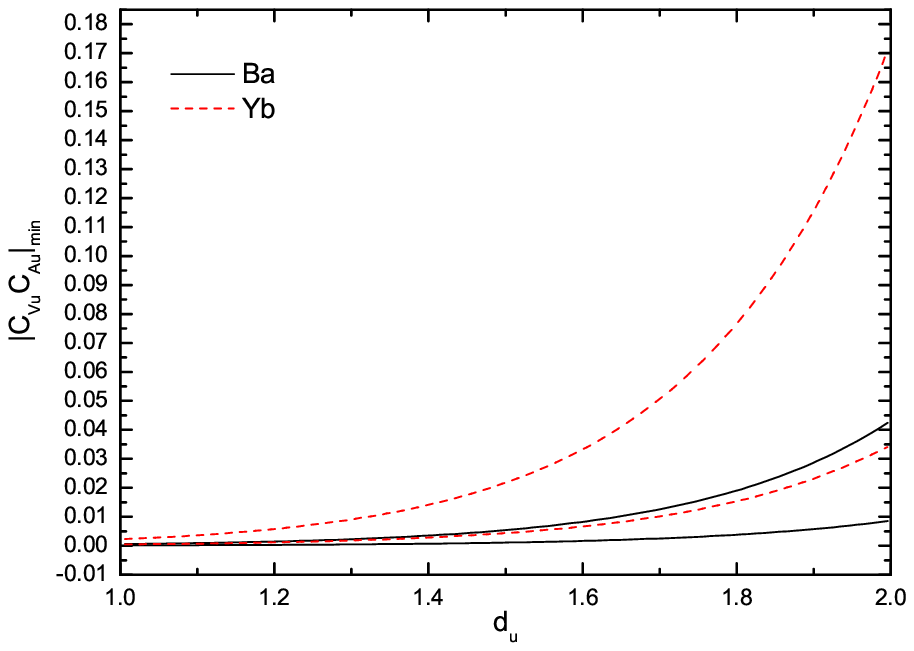}
\caption{\label{fig1} The lower bound on
$|c_{V\cal{U}}\,c_{A\cal{U}}|$ as a function of $d_{\cal U}$ imposed
by extracting unparticle physics from APV along the Ba and Yb
isotope. The two upper curves correspond to the uncertainty
$\delta(\Delta r_N)=1.0\times10^{-4}$, and the two lower curves
correspond to $\delta(\Delta r_N)=5.0\times10^{-4}$. }
\end{center}
\end{figure}

The uncertainties in the neutron distribution induce uncertainties
in the ratios $R_1$ and $R_2$, which is given by
\begin{eqnarray}
\nonumber&&\delta(\delta^{nucl}_1)=-(Z\alpha)^2\frac{N'}{\Delta
N}[f'(r_N)\delta(\Delta r_N)+f''(r_N)\,\delta r_N\Delta r_N]\\
\label{v4eq5}&&\delta(\delta^{nucl}_2)=-(Z\alpha)^2[f'(r_N)\delta(\Delta
r_N)+f''(r_N)\delta r_N\Delta r_N]
\end{eqnarray}
where $\delta r_N$ is the uncertainties in $r_N$. From the view of
extracting the unparticle physics limits, the impact of neutron
distribution uncertainties is characterized by the ratio between
$\delta(\delta^{nucl}_i)$ and the unparticle physics corrections
$\delta^{\cal U}_i(i=1,2)$. The smaller the size of this ratio is,
the less problematic the neutron distribution uncertainties become.
Straightforwardly we have
\begin{eqnarray}
\nonumber&&\delta(\delta^{nucl}_1)/\delta^{\cal
U}_1\simeq-(Z\alpha)^2\frac{N'}{\Delta
N}\frac{N+N'}{2Z}\frac{\Gamma(d_{\cal U})(2d_{\cal
U}-1)}{(2\pi)^{3/2-2d_{\cal U}}\Gamma(d_{\cal
U}+\frac{1}{2})}\frac{f'(r_N)\,\delta(\Delta r_N)\Lambda^2_{\cal
U}\,G_F}{24\,c_{V{\cal U}}c_{A{\cal U}}}\\
\label{v4eq6}&&\delta(\delta^{nucl}_2)/\delta^{\cal
U}_2\simeq-(Z\alpha)^2\frac{N'}{\Delta
N}\frac{N}{Z}\frac{\Gamma(d_{\cal U})(2d_{\cal
U}-1)}{(2\pi)^{3/2-2d_{\cal U}}\Gamma(d_{\cal
U}+\frac{1}{2})}\frac{f'(r_N)\,\delta(\Delta r_N)\Lambda^2_{\cal
U}\,G_F}{24\,c_{V{\cal U}}c_{A{\cal U}}}
\end{eqnarray}
where the terms containing $\delta r_N\Delta r_N$ have been dropped,
since generally one has $\delta r_N\Delta r_N\ll\delta(\Delta
r_N)$\cite{nuclstu}. It is obvious that
$\delta(\delta^{nucl}_1)/\delta^{\cal
U}_1\simeq\delta(\delta^{nucl}_2)/\delta^{\cal U}_2$. Although $R_1$
is more sensitive to unparticle physics by $\frac{N'}{\Delta N}$ as
compared to $R_2$, it is approximately sensitive to the neutron
distribution uncertainties by the same factor as well. In order to
constraint unparticle physics from APV along an isotope, the
uncertainties due to neutron distribution at least should be smaller
than the unparticle contributions. For demonstration, the lower
bound on the absolute value of $c_{V\cal{U}}\,c_{A\cal{U}}$ as a
function of $d_{\cal U}$ for the Ba and Yb isotope are displayed in
Fig.\ref{fig1}.

\section{unparticle and ${\bf\rm{NuTeV}}$ anomaly}

A few years ago, the NuTeV collaboration at Fermilab measured the
value of the Weinberg angle $\sin^2\theta_w$ in deep inelastic
scattering on nuclear target with both neutrino and antineutrino
beams. Having considered and examined various source of systematic
errors, the NuTeV collaboration reported the value
$\sin^2\theta_w=0.2277\pm0.0013(stat)\pm0.0009(syst)$\cite{nutev}.
This result is about $3\sigma$ deviation from the world average
value $\sin^2\theta_w=0.2227\pm0.00037$, which is measured in other
electroweak processes\cite{pdg}. A number of possible solutions to
this deviation have been proposed, such as strange-antistrange
asymmetry\cite{strange}, tiny violation of isospin symmetry in
PDFs\cite{isospin}, nuclear physics effects\cite{nuclear}, the
effects of the next-to-leading order QCD corrections\cite{qcd} and
the new physics beyond the SM. For review, please see
Ref.\cite{review}. So far the NuTeV anomaly is still an open
question. In the following, we will explore the possible effects of
unparticle in NuTeV experiment and the constraints on the unparticle
parameters, if the NuTeV results are confirmed in future.

The NuTeV collaboration measured the value of $\sin^2\theta_w$ by
using the ratios of the neutral-current to charged-current total
cross section on iron for neutrino and antineutrino respectively,
this procedure is closely related the Paschos-Wolfenstein(PW)
relation\cite{pw}
\begin{equation}
\label{nutev1}R^{-}=\frac{\sigma^{\nu
N}_{NC}-\sigma^{\bar{\nu}N}_{NC}}{\sigma^{\nu
N}_{CC}-\sigma^{\bar{\nu}N}_{CC}}=\frac{1}{2}-\sin^2\theta_w
\end{equation}
where $\sigma^{\nu N}_{CC}$ and $\sigma^{\nu N}_{NC}$ are
respectively the charged-current(CC) and neutral-current(NC) cross
sections in neutrino-nucleon deep inelastic scattering processes,
and $\sigma^{\bar{\nu} N}_{CC}$ and $\sigma^{\bar{\nu} N}_{NC}$ are
corresponding antineutrino cross sections. The relationship
Eq.(\ref{nutev1}) is based on the assumptions of isoscalar target
and strange-antistrange symmetry of the nuclear sea. Later the
corrections from unparticle physics and strange quark asymmetry
would be included, nuclear physics effects and non-isoscalar
corrections etc are discussed as well.

In Ref.\cite{ding} we have demonstrated that the cross sections for
neutrino- and antineutrino-nucleon neutral current interactions have
the following form:
\begin{equation}
\label{3}\frac{d^2\sigma^{\nu(\bar{\nu})A}_{NC}}{dxdy}=s\;\frac{G^2_F}{2\pi
}(\frac{M^2_Z}{Q^2+M^2_Z})^2[\;xy^2F^{A,NC}_1+(1-y)F^{A,NC}_2\pm
y(1-\frac{1}{2}y)xF^{A,NC}_3\;]
\end{equation}
where $G_F$ is the Fermi constant
$G_F\simeq1.166\times10^{-5}\rm{GeV}^2$\cite{pdg}, $Q^2=-q^2$,
$x=Q^2/2p\cdot q$, $y=p\cdot q/p\cdot k$ and $s=(k+p)^2$ are the
standard Bjorken DIS variables for the four momentum $k$($p$) of the
initial state neutrino or antineutrino(nucleon), and $A$ denotes the
target. At the leading order, the structure functions
$F^{A,NC}_i(x,Q^2)$(i=1,2,3) are expressed in terms of the quark and
antiquark distributions as follows,
\begin{eqnarray}
\nonumber F^{A,NC}_1(x,Q^2)&=&\sum_{q}\;[q^{A}(x)+\overline{q}^{A}(x)]B^{\nu}_q(Q^2)\\
\nonumber
F^{A,NC}_2(x,Q^2)&=&\sum_{q}x[q^{A}(x)+\overline{q}^{A}(x)]C^{\nu}_q(Q^2)\\
\label{4}
F^{A,NC}_3(x,Q^2)&=&\sum_{q}[q^{A}(x)-\overline{q}^{A}(x)]D^{\nu}_q(Q^2)
\end{eqnarray}
with
\begin{eqnarray}
\nonumber B^{\nu}_q(Q^2)&=&\frac{V^2_q+A^2_q}{2}+\frac{(c^2_{V{\cal
U}}+c^2_{A{\cal U}})^2}{2G^2_FM^{4}_{Z}\Lambda^{4d_{\cal U}-4}_{\cal
U}}R^2_{{\cal U}Z}+\frac{(V_q\;c_{V{\cal U}}-A_q\;c_{A{\cal
U}})(c_{V{\cal
U}}-c_{A{\cal U}})}{\sqrt{2}\;G_FM^{2}_Z\Lambda^{2d_{\cal U}-2}_{\cal U}}R_{{\cal U}Z}\\
\label{5}&&+\frac{(c^2_{S{\cal U}}-c^2_{P{\cal
U}})^2}{4G^2_FM^{4}_{Z}\Lambda^{4d_{\cal U}-4}_{\cal U}}R^2_{{\cal U}Z}\\
\label{6}C^{\nu}_q(Q^2)&=&V^2_q+A^2_q+\frac{(c^2_{V{\cal
U}}+c^2_{A{\cal U}})^2}{G^2_FM^{4}_Z\Lambda^{4d_{\cal U}-4}_{\cal
U}}R^2_{{\cal U}Z}+\frac{\sqrt{2}\,(V_q\;c_{V{\cal
U}}-A_q\;c_{A{\cal U}})(c_{V{\cal
U}}-c_{A{\cal U}})}{G_FM^2_Z\Lambda^{2d_{\cal U}-2}_{\cal U}}R_{{\cal U}Z}\\
\label{7}D^{\nu}_q(Q^2)&=&2V_qA_{q}+\frac{4\,c^2_{V{\cal
U}}\;c^2_{A{\cal U}}}{G^2_FM^{4}_Z\Lambda^{4d_{\cal U}-4}_{\cal
U}}R^2_{{\cal U}Z}-\frac{\sqrt{2}\;(V_q\;c_{A{\cal
U}}-A_q\;c_{V{\cal U}})(c_{V{\cal
U}}-c_{A{\cal U}})}{G_FM^2_{Z}\Lambda^{2d_{\cal U}-2}_{\cal U}}R_{{\cal U}Z}\\
\label{8}R_{{\cal U}Z}&=&\frac{A_{d_{\cal U}}}{2\sin(d_{\cal
U}\pi)}(Q^2)^{d_{\cal U}-2}(Q^2+M^2_Z)
\end{eqnarray}
here $V_q=T_{3q}-2Q_{q}\sin^2\theta_W$, $A_{q}=T_{3q}$, $Q_q$ and
$T_{3q}$ are respectively the electric charge and the third
component of the weak isospin of the quark $q$, and $A_{d_{\cal
U}}=\frac{16\pi^{5/2}}{(2\pi)^{2d_{\cal U}}}\frac{\Gamma(d_{\cal
U}+1/2)}{\Gamma(d_{\cal U}-1)\Gamma(2d_{\cal U})}$, which is the
normalization factor of the state density for the unparticle
stuff\cite{georgi1}.

The neutrino- and antineutrino-nucleus charged current cross
sections are expressed in a similar manner:
\begin{equation}
\label{9}\frac{d^2\sigma^{\nu(\bar{\nu})A}_{CC}}{dxdy}=s\;\frac{G^2_F}{2\pi
}(\frac{M^2_W}{Q^2+M^2_W})^2[\;xy^2F^{\nu(\bar{\nu})A,CC}_1+(1-y)F^{\nu(\bar{\nu})A,CC}_2\pm
y(1-\frac{1}{2}y)xF^{\nu(\bar{\nu})A,CC}_3\;]
\end{equation}
The above charge-current structure functions are given by:
\begin{eqnarray}
\label{10}F^{\nu
A,CC}_1(x,Q^2)&=&\sum_{i,j}[d^{A}_{j}(x)+\bar{u}^{A}_i(x)]|(V_{CKM})_{ij}|^2\\
\label{11}F^{\nu A,CC}_2(x,Q^2)&=&2xF^{\nu A,CC}_1(x,Q^2)\\
\label{12}F^{\nu A,CC}_3(x,Q^2)&=&2\sum_{ij}[d^{A}_{j}(x)-\bar{u}^{A}_i(x)]|(V_{CKM})_{ij}|^2\\
\label{13}F^{\bar{\nu}
A,CC}_1(x,Q^2)&=&\sum_{i,j}[u^{A}_{i}(x)+\bar{d}^{A}_j(x)]|(V_{CKM})_{ij}|^2\\
\label{14}F^{\bar{\nu}A,CC}_2(x,Q^2)&=&2xF^{\bar{\nu}A,CC}_1(x,Q^2)\\
\label{15}F^{\bar{\nu}
A,CC}_3(x,Q^2)&=&2\sum_{ij}[u^{A}_{i}(x)-\bar{d}^{A}_j(x)]|(V_{CKM})_{ij}|^2
\end{eqnarray}
where $V_{CKM}$ is the Cabibbo-Kobayashi-Maskawa matrix, and $i,j$
is the generation index. In the NuTeV experiment, the average value
of $Q^2$ is about 20$\rm{GeV^2}$($Q^2\ll M^2_Z,M^2_W$). Substituting
Eqs.(\ref{3}-\ref{15}) into the right hand of Eq.(\ref{nutev1}), to
leading order of the small coupling constants $c_{V{\cal U}}$,
$c_{A{\cal U}}$, $c_{S{\cal U}}$ and $c_{P{\cal U}}$, we obtain the
modified Paschos-Wolfenstein(PW) relation:
\begin{equation}
\label{16}R^{-}=\frac{1}{2}-\sin^2\theta_w-\delta R^{-}
\end{equation}
with the correction $\delta R^{-}$ is
\begin{eqnarray}
\nonumber\delta
R^{-}\simeq&&\Big\{(1-\frac{7}{3}\sin^2\theta_w)S_v+\frac{\sqrt{2}R_{{\cal
U}Z}(c_{V{\cal U}}-c_{A{\cal U}})}{G_FM^2_{Z}\Lambda_{\cal
U}^{2d_{\cal U}-2}}\Big(-\frac{1}{3}\sin^2\theta_w\;c_{A{\cal
U}}\;Q_v+[(-\frac{1}{2}+\frac{2}{3}\sin^2\theta_w)\;c_{A{\cal
U}}\\
\label{17}&&+\frac{1}{2}c_{V{\cal U}}]S_v\Big)\Big\}/(Q_v+3S_v)
\end{eqnarray}
where
$Q_v=\int^{1}_0x[u^{p}(x)+d^{p}(x)-\bar{u}^{p}(x)-\bar{d}^{p}(x)]dx$
and $S_v=\int^{1}_0x[s^{p}(x)-\bar{s}^{p}(x)]dx$. Since a small
strange-antistrange asymmetry could be responsible for a significant
fraction of the observed discrepancy, we have explicitly included
this asymmetry in our analysis. From the analysis of the parton
distributions\cite{fit2}, we choose $Q_v\approx0.36$ following the
authors in Ref.\cite{review}, which is better than 10\% accuracy in
the energy range of the NuTeV experiment. While the situation
about$S_v$ is not so clear until now. In Ref.\cite{fit1} a global
fit to all available neutrino data found evidence in favor of the
strange sea asymmetry. The CTEQ group \cite{Olness:2003wz} has
performed a global QCD analysis by including the dimuon data from
the CCFR and NuTeV collaboration\cite{dilepton}, they found that a
large negative $S_v$ is strongly disfavored by both dimuon and other
inclusive data, the strange asymmetry was in the range
$-0.001<S_v<0.004$ and the most likely value was $S_v\sim0.002$.
Theoretically, phenomenologocal model calculations suggest
$S_v\approx0.002$\cite{Brodsky:1996hc} , which seems compatible with
all the present experimental information. Since we can not
unambiguously fix the asymmetry parameter $S_v$ from both the global
data fit and theoretical calculations, the $S_v$ dependence of
$\delta R^{-}$ would be illustrated later.

\begin{figure}[hptb]
\centering
\begin{minipage}[t]{0.48\textwidth}
\centering
\includegraphics[width=9cm]{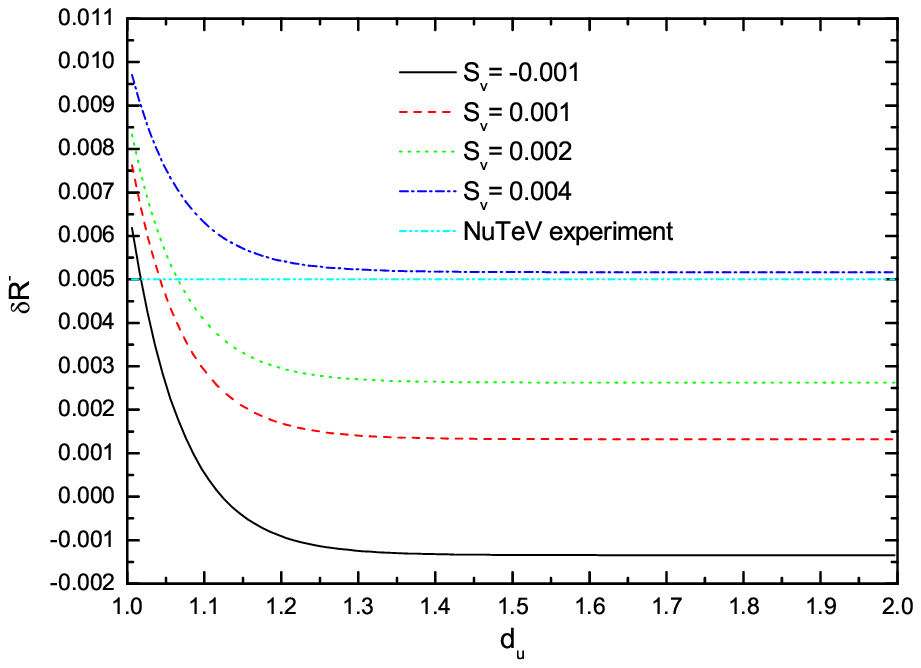}
\caption{\label{fig3} The dependence of $\delta R^{-}$ on $d_{\cal
U}$ for $S_v=$-0.001, 0.001, 0.002 and 0.004 respectively with
$\Lambda_{\cal U}=$1TeV, $c_{V{\cal U}}=0.01$ and $c_{A{\cal
U}}=0.002$.}
\end{minipage}%
\hspace{0.04\textwidth}%
\begin{minipage}[t]{0.48\textwidth}
\centering
\includegraphics[width=9cm]{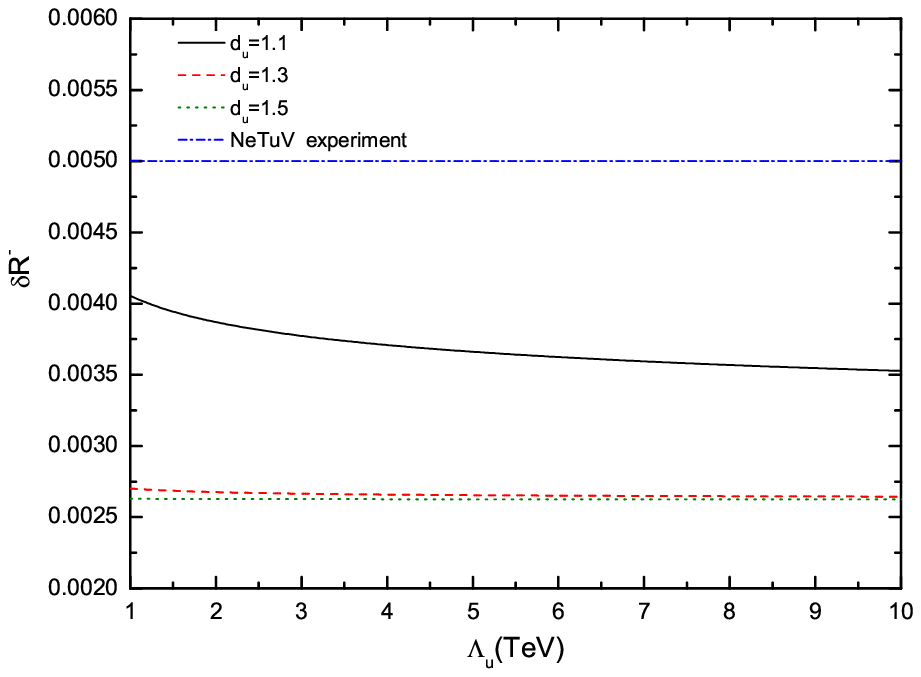}
\caption{\label{fig4} The dependence of $\delta R^{-}$ on
$\Lambda_{\cal U}$ for $d_{\cal U}=$1.1, 1.3 and 1.5 with
$\Lambda_{\cal U}=$1TeV, $c_{V{\cal U}}=0.01$ and $c_{A{\cal
U}}=0.002$.}
\end{minipage}
\end{figure}

As is shown in Eq.(\ref{17}), the correction to the PW relation
induced by unparticle and the strange-antistrange asymmetry
sensitively depends on the scale dimension $d_{\cal U}$ and the
unparticle scale $\Lambda_{\cal U}$. For illustration, we display
$\delta R^{-}$ as a function of $d_{\cal U}$ for $S_v=$-0.001,
0.001, 0.002 and 0.004 respectively in Fig.\ref{fig3}. The
variations of $\delta R^{-}$ with respect to $\Lambda_{\cal U}$ are
shown in Fig.\ref{fig4}. From Fig.\ref{fig3} and Fig.\ref{fig4}, we
see that the PW relation receives large corrections from unparticle
for $d_{\cal U}$ near 1.

To get an idea of the possible role played by unparticle in
dissolving NuTeV anomaly, we first discuss four special cases in the
following.
\begin{enumerate}
\item {No unparticle physics $c_{V{\cal U}}=c_{A{\cal U}}=0$}

In this case the correction $\delta R^{-}$ in Eq.(\ref{17}) reduces
to $(1-\frac{7}{3}\sin^2\theta_w)\frac{S_v}{Q_v+3S_v}$, which is
exactly the correction to the PW relation due to the asymmetric
strange sea\cite{strange}.
\item{Pure vector coupling between
unparticle and SM fermion $c_{V{\cal U}}\neq0$ and $c_{A{\cal
U}}=0$}

The corresponding correction to the PW relation is denoted as
$\delta R^{-}_V$, then
\begin{equation}
\label{18}\delta
R^{-}_V=[1-\frac{7}{3}\sin^2\theta_w+\frac{\sqrt{2}\,R_{{\cal
U}Z}c^2_{V{\cal U}}}{2G_FM^2_Z\Lambda_{\cal U}^{2d_{\cal
U}-2}}]\frac{S_v}{Q_v+3S_v}
\end{equation}
For $1<d_{\cal U}<2$, it is obvious that $R_{{\cal U}Z}$ is
negative, so that the unparticle contributions tend to cancels the
contribution from the strange quark asymmetry. For the most likely
value $S_v\simeq0.002$,  unparticle would increase the discrepancy
between the NuTeV $\sin^2\theta_w$ result and its SM value in this
case.
\item{Pure axial vector coupling between
unparticle and SM fermion $c_{A{\cal U}}\neq0$ and $c_{V{\cal
U}}=0$}

The correction to the PW relation is then denoted as $\delta
R^{-}_A$, which can be straightforwardly obtained from Eq.(\ref{17})
\begin{equation}
\label{19}\delta
R^{-}_A=\Big\{1-\frac{7}{3}\sin^2\theta_w+\frac{\sqrt{2}\,R_{{\cal
U}Z}c^2_{A{\cal U}}}{G_FM^2_{Z}\Lambda_{\cal U}^{2d_{\cal
U}-2}}\;[\frac{1}{2}-\frac{2}{3}\sin^2\theta_w+\frac{\sin^2\theta_w}{3}\;\frac{Q_v}{S_v}]\Big\}\frac{S_v}{(Q_v+3S_v)}
\end{equation}
For convenience, we define the factor
$F\equiv\frac{1}{2}-\frac{2}{3}\sin^2\theta_w+\frac{\sin^2\theta_w}{3}\;\frac{Q_v}{S_v}$,
which determines the sign of unparticle contributions relative to
that of the strange quark asymmetry. If the strange asymmetry $S_v$
is negative(i.e. $-0.001<S_v<0$), the factor $F$ is negative, the
contributions of unparticle increase those of asymmetric strange
quark. Whereas for positive $S_v$ (i.e. $0<S_v<0.004$), $F$ is
positive, consequently the signs of the contributions from
unparticle and the strange quark asymmetry are opposite. For the
central value $S_v\simeq0.002$, the NuTeV discrepancy would be
increase by unparticle as well.
\item{Chiral symmetric
coupling limit $c_{V{\cal U}}=c_{A{\cal U}}$}

From Eq.(\ref{17}), it is obvious that there is no correction to the
PW relation from unparticle exchange, then $\delta R^{-}$ completely
comes from strange quark asymmetry.
\end{enumerate}
Generally the unparticle parameters $\Lambda_{\cal U}$, $d_{\cal
U}$, $c_{V{\cal U}}$ and $c_{A{\cal U}}$ could take value beyond the
four special cases considered above. We are very interested in
exploring if there exists certain parameter region, where the NuTeV
discrepancy is completely accounted for by the unparticle effects
and the asymmetric strange-antistrange sea. If this is true, the
following relation should be satisfied
\begin{eqnarray}
\nonumber&&\frac{\sqrt{2}\,R_{{\cal U}Z}(c_{V{\cal U}}-c_{A{\cal
U}})}{G_FM^2_{Z}\Lambda_{\cal U}^{2d_{\cal
U}-2}}\{\frac{1}{2}S_vc_{V{\cal U}}+[-\frac{1}{3}\sin^2\theta_wQ_v
+(-\frac{1}{2}+\frac{2}{3}\sin^2\theta_w)S_v]c_{A{\cal U}}\}\\
\label{20}&&= 5\times
10^{-3}(Q_v+3S_v)-(1-\frac{7}{3}\sin^2\theta_w)S_v
\end{eqnarray}

\begin{figure}[hptb]
\begin{center}
\begin{tabular}{cc}
\scalebox{0.5}{\includegraphics{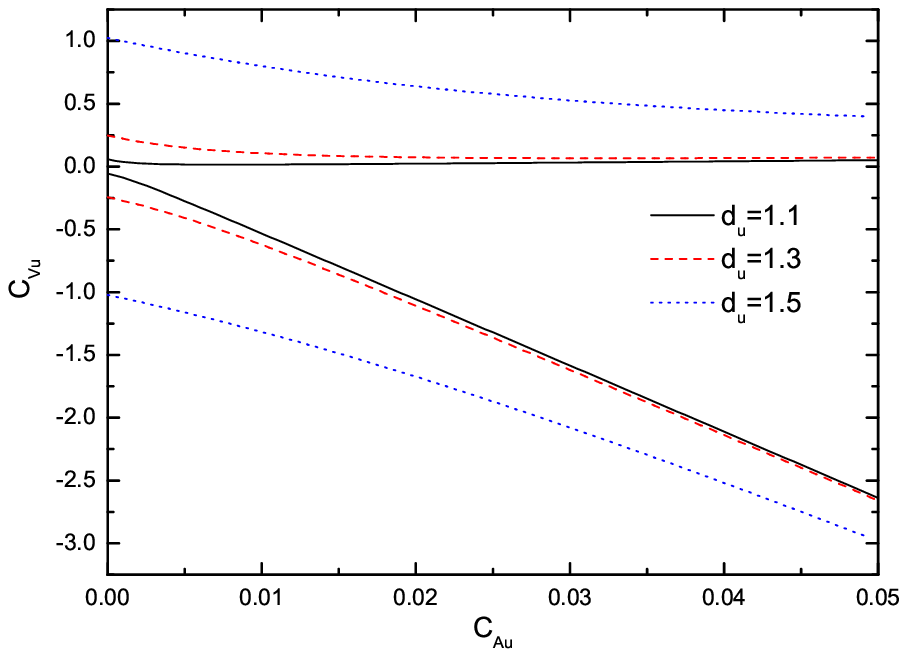}}&
\scalebox{0.5}{\includegraphics{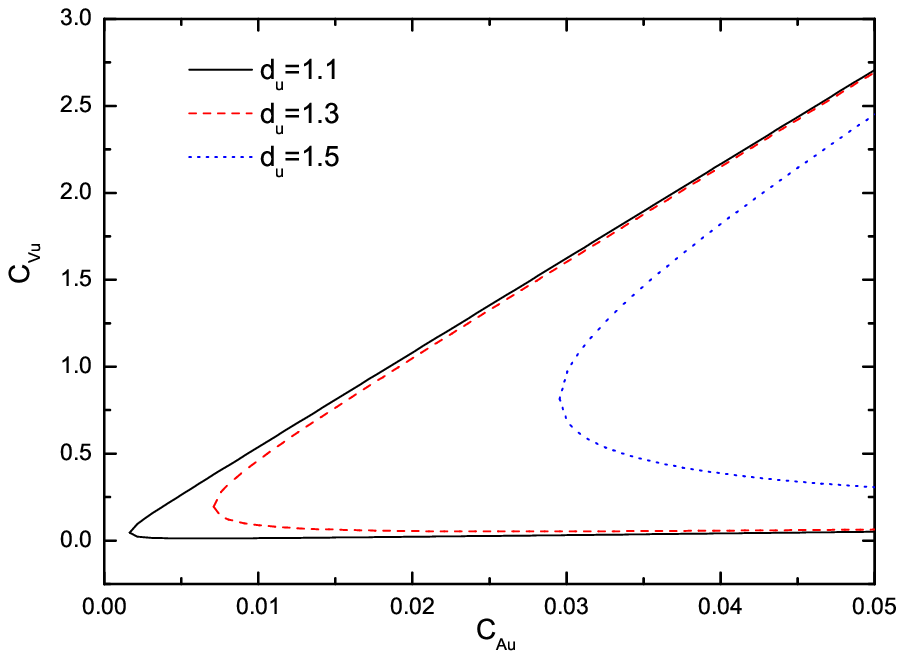}} \\
(a)&(b)\\
\scalebox{0.5}{\includegraphics{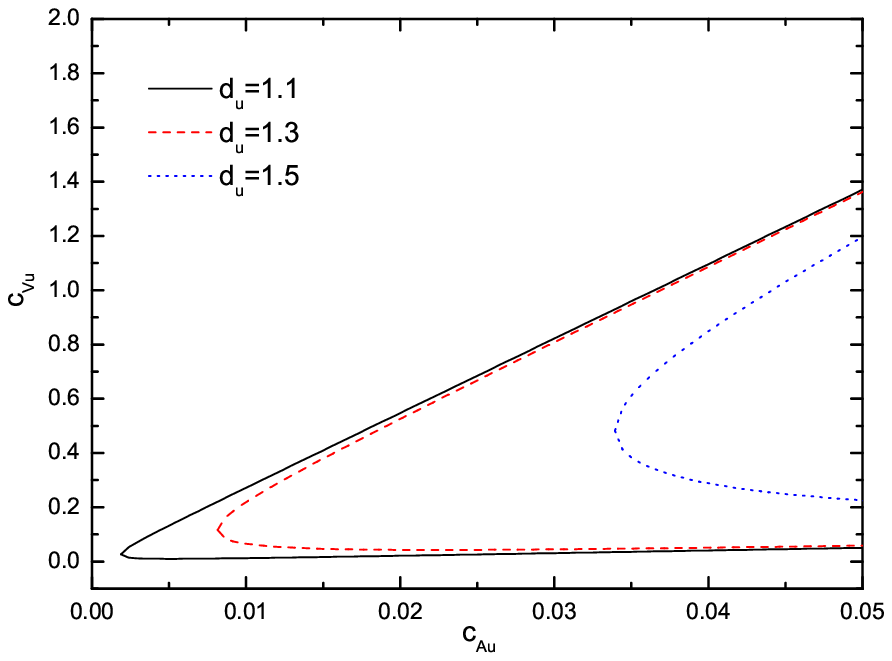}}&
\scalebox{0.5}{\includegraphics{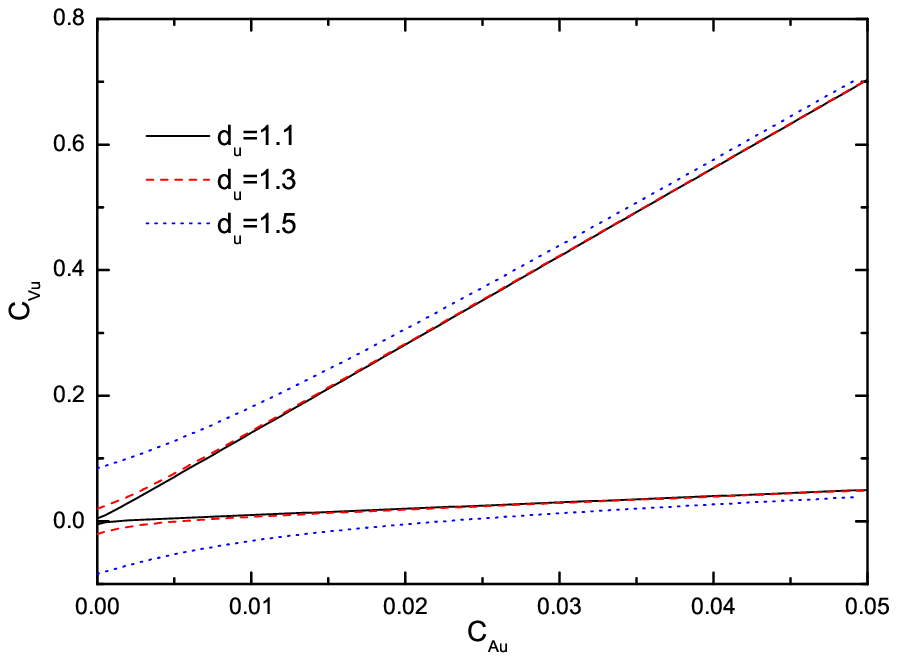}}\\
(c)&(d)
\end{tabular}
\caption{\label{fig2} The curves of $c_{V{\cal U}}$ with respect to
$c_{A{\cal U}}$ for $d_{\cal U}=1.1, 1.3, 1.5$, and contributions to
$\delta R^-$ from isoscalarity violation etc are assumed to be zero.
Fig.\ref{fig2}(a), Fig.\ref{fig2}(b), Fig.\ref{fig2}(c) and
Fig.\ref{fig2}(d) correspond to $S_v=-0.001$, 0.001, 0.002 and 0.004
respectively.}
\end{center}
\end{figure}
The above relation between $c_{V{\cal U}}$ and $c_{A{\cal U}}$ for
$d_{\cal U}=1.1, 1.3, 1.5$ are displayed in Fig.\ref{fig2}, where we
have chosen $\Lambda_{\cal U}=1$ TeV and $Q^2\approx20\rm{GeV}^2$.
To demonstrate the $S_v$ dependence, the parameter regions for
$S_v=-0.001,\,0.001,\,0.002,\, {\rm and}\, 0.004$ are shown
respectively. In the case of $S_v>0$, the contributions to $\delta
R^{-}$ from the unparticle physics and asymmetric strange quark are
larger than the NuTeV discrepancy 0.005 for the parameters lying in
the region between the upper curve and the lower curve of the same
type(solid line, dashed line or dotted line). Whereas the
contributions are smaller than the NuTeV discrepancy for the
parameters in the region above the upper curve and that below the
lower curve of the same type. However, the reverse is true in the
case of $S_v<0$. Particularly we note that the NuTeV anomaly is
completely dissolved by the combination of unparticle effect and
strange quark asymmetry, for the parameters on the curves.

In general, the effective unparticle couplings $c_{V{\cal U}}$ and
$c_{A{\cal U}}$ are expected to be small, some literatures of
unparticle physics have chosen them to be of order $10^{-2}$ or
$10^{-3}$. From Fig.\ref{fig2} we see that, for the parameters on
the upper curves with $S_v<0$ and the parameters on the lower curves
with $S_v>0$, $c_{V{\cal U}}$ and $c_{A{\cal U}}$ can take values in
the range  from $10^{-3}$ to $10^{-2}$, and simultaneously the NuTeV
anomaly can be explained by unparticle effects and strange quark
asymmetry. The corresponding parameter regions for $d_{\cal U}=1.1$,
1.3 and 1.5 are shown in Fig.\ref{region1}, where the variation of
$S_v$ in the range $-0.001<S_v<0.004$ is considered. Looking back at
Fig.{\ref{fig1}}, we can see that for these parameter regions, the
unparticle effects may be observed in APV, if the uncertainty in the
isotopic relative neutron/proton radius shift
$\delta(\Delta\frac{R_N}{R_P})$ is smaller than a few times
$10^{-4}$. Moreover, it is notable that these parameter spaces are
consistent with the constraints on unparticle parameters from
$b\rightarrow s\gamma$\cite{He:2008xv}.

The NuTeV experiment uses an iron target, which has an excess of
neutrons over protons about $6\%$. This resulting isoscalarity
violation would introduce correction to the PW
relation\cite{Kulagin:2003wz}. This violation of isoscalarity is
known to good accuracy, it has been claimed to be included in the
data analysis by the NuTeV collaboration\cite{Zeller:2002du}. A
further violation of isoscalarity could be due to the fact that
isospin symmetry is violated by the parton distributions of the
nucleon, i.e. $u^{p}\neq d^n$ and $u^n\neq d^p$, where $u^p$ and
$d^p$ are respectively the up quark and down quark distributions in
proton, $u^n$ and $d^n$ are the corresponding distributions in
neutron. The contribution of this type of isospin violation (it is
named as charge symmetry violation in Ref.\cite{isospin}) is
expected to reduce the discrepancy of $\sin^2\theta_W$ by about
$30\%$\cite{Zeller:2002du,Sather:1991je,Rodionov:1994cg}.

\begin{figure}[hptb]
\begin{center}
\begin{tabular}{cc}
\scalebox{0.5}{\includegraphics{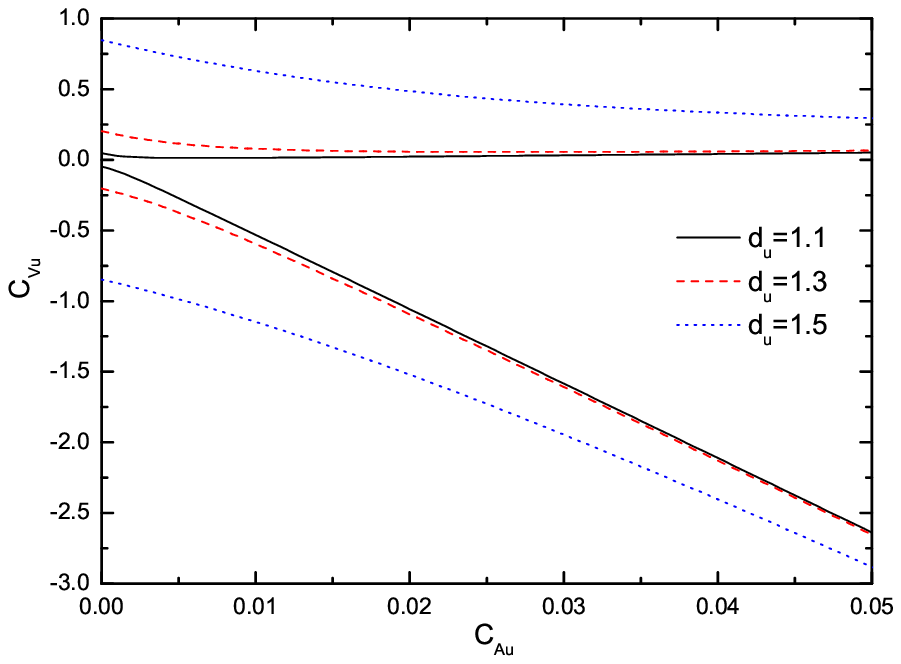}}&
\scalebox{0.5}{\includegraphics{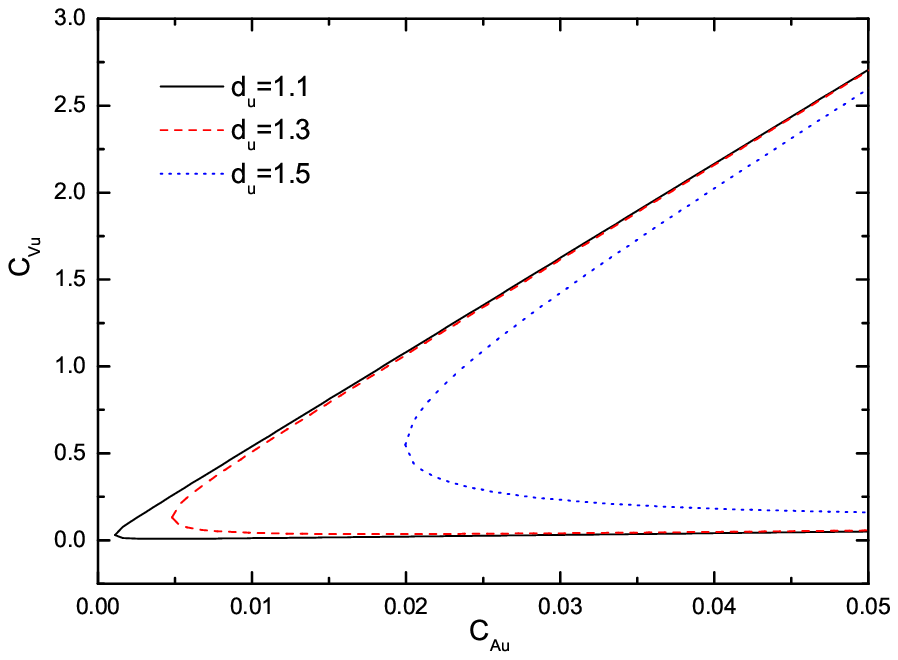}} \\
(a)&(b)\\
\scalebox{0.5}{\includegraphics{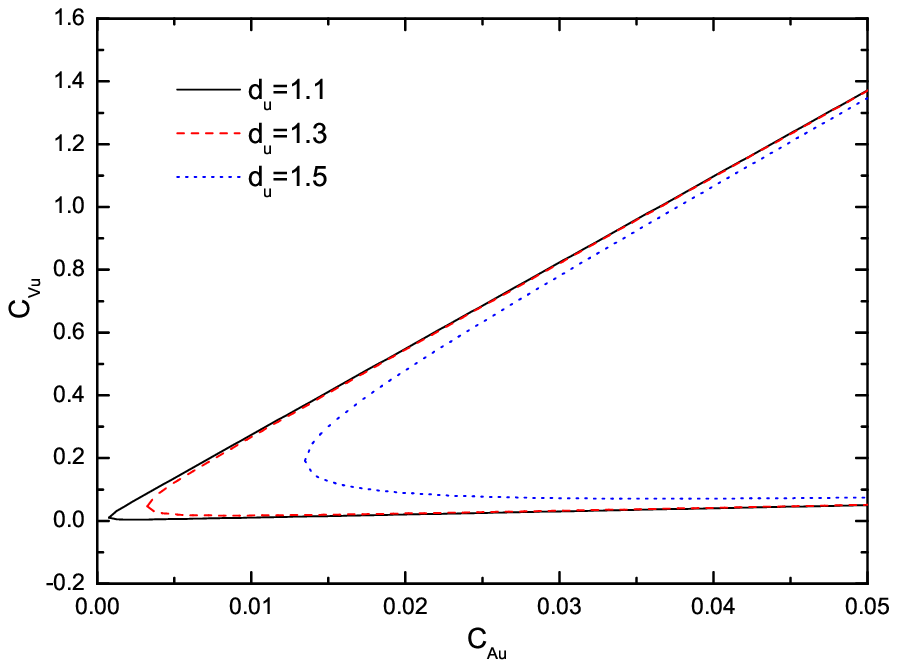}}&
\scalebox{0.5}{\includegraphics{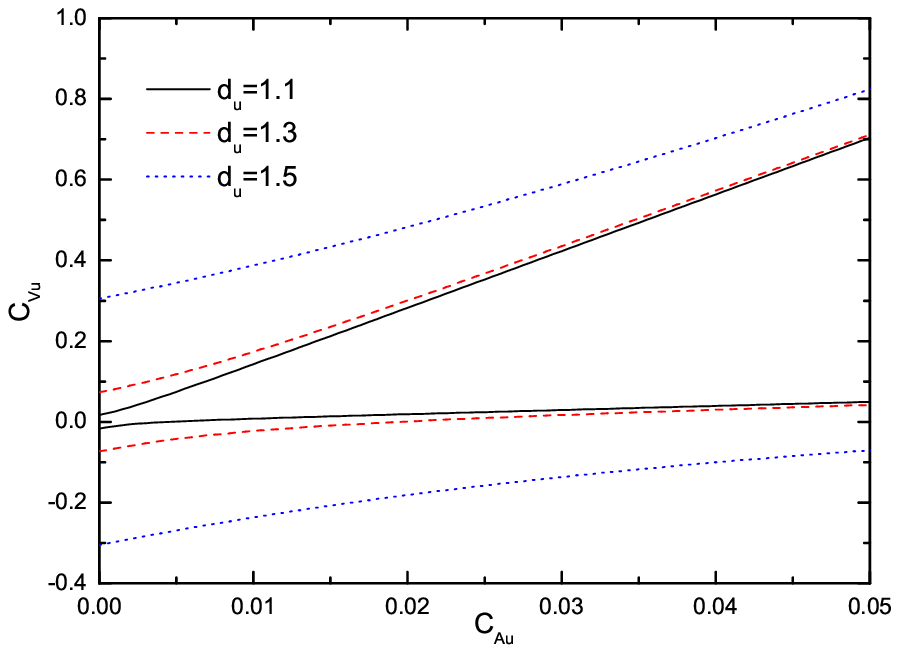}}\\
(c)&(d)
\end{tabular}
\caption{\label{addfig}The same as Fig.\ref{fig2}, and the
contributions to $\delta R^-$ from isoscalarity violation etc are
taken to be 0.002. }
\end{center}
\end{figure}

The PW relation may receive corrections from higher order QCD
contributions and nuclear physics effects (such as Fermi motion,
nuclear binding and nuclear shadowing etc) as well, which turn out
to be small enough to be
negligible\cite{Kulagin:2003wz,Kulagin:2004xs}. In short, there are
theoretical corrections on the NuTeV determination of
$\sin^2\theta_W$ due to QCD effects, nuclear effects and the
violation of the assumptions on which the PW relation is based. The
calculations of some corrections are model-dependent. For
demonstration, we assume that the contributions to $\delta R^-$ from
isoscalarity violation etc are 0.002, then the corresponding
relationship between $c_{V{\cal U}}$ and $c_{A{\cal U}}$ are shown
in Fig.\ref{addfig}. The meaning of these figure is the same as that
of Fig.\ref{fig2}, and the shapes of the curves in the two figures
are similar to each other. Especially for the parameter values on
the upper curves with $S_v<0$ and the parameter values on the lower
curves with $S_v>0$, the unparticle couplings $c_{V{\cal U}}$ and
$c_{A{\cal U}}$ can take values of order from $10^{-3}$ to
$10^{-2}$, which are also consistent with the constraints on
unparticle parameters from $b\rightarrow s\gamma$\cite{He:2008xv}.
At the same time, the residual NuTeV anomaly is completely removed
by the unparticle physics and the asymmetric strange quark
contribution. Explicitly, the corresponding parameter regions are
shown in Fig.\ref{region2}.

\section{summary and conclusions}

Unparticle leads to interesting and rich phenomenology, which could
be checked by experiments. In this work we have investigated the
possible signals of unparticle in APV along an isotope chain and
NuTeV experiment. Both APV and NuTeV experiment would play important
role in exploring unparticle physics at lower energy, if unparticle
exists in nature. Possible theoretical corrections and uncertainties
within the standard model are discussed in detail. In the case of
the ratios of APV observables along an isotope chain, $R_1$ is more
sensitive to unparticle than $R_2$ provided the same experimental
precisions. Although the theoretical uncertainties from atomic
theory almost cancel, the uncertainties due to neutron distribution
are crucial. The effects of unparticle could be observed, if the
uncertainty in relative neutron/proton radius shift
$\delta(\Delta\frac{R_N}{R_P})$ is less than a few times $10^{-4}$
by measuring the parity violating electron scattering.

The interpretation of the NuTeV result is still a subject of
considerable debate until now. Various effects unaccounted for by
the NuTeV collaboration, have been proposed as possible remedies for
the anomaly. The NuTeV results should be checked by other groups or
other experiments such as Q-Weak experiment in future, if the NuTeV
discrepancy is confirmed to be true, we suggest that unparticle
could remove part of the discrepancy. The constraints imposed by
NuTeV experiment on unparticle physics are studied in detail. We
have demonstrated that there exist certain regions for the
unparticle parameters ($\Lambda_{\cal U}$, $d_{\cal U}$, $c_{V{\cal
U}}$ and $c_{A{\cal U}}$), where the unparticle physics coupled with
the strange quark asymmetry could completely explain the discrepancy
between the NuTeV result and the SM value with or without the
contributions from the isoscalarity violation etc. It is remarkable
that these parameter regions consistent with the constraints from
$b\rightarrow s\gamma$\cite{He:2008xv}. Meanwhile, unparticle
possibly manifests itself in APV for these parameter values, if the
netron distribution uncertaintie $\delta(\Delta\frac{R_N}{R_P})$ is
less than a few times $10^{-4}$.

Both the APV along the isotope chain and the NuTeV results put
important constraints on the unparticle physics. The unparticle
contributions to these observables strongly depend on the unparticle
scale $\Lambda_{\cal U}$ and the scale dimension $d_{\cal U}$. It is
very interesting and valuable to perform a global fit to the ranges
of unparticle parameters $\Lambda_{\cal U}$, $d_{\cal U}$,
$c_{V{\cal U}}$, $c_{A{\cal U}}$ allowed by the current precise
electroweak data and other observables from astrophysics and
cosmology etc, which is beyond the scope of the present work.

\section *{ACKNOWLEDGEMENTS}

\indent This work is partially supported by National Natural Science
Foundation of China under Grant Numbers 90403021, and KJCX2-SW-N10
of the Chinese Academy.

\newpage

\begin{figure}[hptb]
\begin{center}
\begin{tabular}{ccc}
(a)&~~~~~~~~~~~~~~~(b)&~~~~~~~~~~~~~~~(c)
\end{tabular}
\caption{\label{region1}The allowed parameter regions of the
effective unparticle couplings $c_{V{\cal U}}$ and $c_{A{\cal U}}$
for $-0.001<S_v<0.004$, where the NuTeV discrepancy are completely
accounted for by unparticle effect and the asymmetric strange sea.
The contributions to $\delta R^-$ from isoscalarity violation etc
are assumed to be zero. Fig.\ref{region1}(a), Fig.\ref{region1}(b)
and Fig.\ref{region1}(c) correspond to $d_u=1.1$, 1.3 and 1.5
respectively.}
\end{center}
\end{figure}

\begin{figure}[hptb]
\begin{center}
\begin{tabular}{ccc}
(a)&~~~~~~~~~~~~~~~(b)&~~~~~~~~~~~~~~~(c)
\end{tabular}
\caption{\label{region2}The same as for Fig.\ref{region1}, and the
contributions to $\delta R^-$ from isoscalarity violation etc are
taken to be 0.002.}
\end{center}
\end{figure}

\end{document}